\documentclass[abbrv,aps,prb,twocolumn,showpacs,preprintnumbers,amsmath,amssymb,
superscriptaddress,nofootinbib]{revtex4}

\usepackage{amsmath,amssymb,amsfonts}
\usepackage{bm,graphicx,color}

\newcommand{\laco}[1]{$\mathrm{LaCoO_3 }$}

\def\lsim{~\rlap{$<$}{\lower 1.0ex\hbox{$\sim$}}}
\def\gsim{~\rlap{$>$}{\lower 1.0ex\hbox{$\sim$}}}

\begin{document}
\title{Excitonic Instability at the Spin-State Transition
	in the Two-Band Hubbard Model}
\author{Jan Kune\v{s} and Pavel Augustinsk\'{y}}
\affiliation{Institute of Physics, Academy of Sciences of the Czech republic, Cukrovarnick\'a 10,
Praha 6, 162 53, Czech Republic}
\date{\today}

\begin{abstract}
Using linear response theory with the dynamical mean-field approximation
we investigate the particle-hole instabilities of the two-band Hubbard
model in the vicinity of the spin-state transition. Besides the previously
reported high-spin--low-spin order we find an instability towards
triplet excitonic condensate. We discuss the strong and weak coupling
limits of the model, in particular, a connection to the spinful hard-core
bosons with a nearest-neighbor interaction. Possible realization
in LaCoO$_3$ at intermediate temperatures is briefly discussed.

\end{abstract}
\maketitle

\section{Introduction}
Search for new states of matter is one of the
central topics of condensed matter physics. While
the development of cold atom techniques allowed
the construction of many exotic phases in particular
in systems of interacting bosons,
electronic order parameters
other than spin, charge and orbital densities
or s-wave pairing superconductivity are rather
rare in real materials.
We report observation of an off-diagonal order close to the spin-state transition
in the two-band Hubbard model with Hund's coupling
and show that such electronic system provides
realization of some of the phases observed
with interacting bosons.

The role of Hund's coupling in correlated electron systems
has been recently theoretically studied in the context of Hund's
metals~\cite{georges13,yin11} and the spin-state
transitions driven by pressure~\cite{kunes08,kunes09} as
well as temperature~\cite{krapek12,eder10} or doping~\cite{august13}.
Competition of different spin states was also linked to the peculiar
magnetic properties of iron pnictides~\cite{chaloupka13}.
The two-band Hubbard model at half filling provides
a minimal lattice realization of the spin-state transition
in correlated electron systems~\cite{werner07,suzuki09}.
Recently, a reentrant transition of Ising type to
a two-sublattice order of high-spin (HS) and low-spin (LS) states
was reported on a bipartite lattice
in the vicinity of the spin-state transition~\cite{kunes11}.
It was proposed that such ordered state can explain
properties of the notorious spin-state transition
compound LaCoO$_3$ at intermediate temperatures.

In this article, we report a systematic investigation
of the particle-hole instabilities in the normal phase of the two-band Hubbard model.
Besides the previously reported Ising instability
we find that an excitonic instability which breaks a continuous symmetry
dominates over a broad range of parameters.
The idea of an instability due to the long-range part
of the Coulomb interaction in small gap semiconductors
leading to so the called excitonic insulator
phase appeared fifty years ago~\cite{halperin68} and more recently
was applied to the physics of LaB$_{6}$~\cite{balents00}.
Following the work of Batista~\cite{batista02} on electronic
ferroelectricity, the excitonic instability was studied
in the extended Falicov-Kimball model~\cite{zenker12,zenker11,seki11}
as well as the two-band Hubbard model without Hund's coupling~\cite{zocher11,kaneko12}.
\begin{figure}
    \includegraphics[height=\columnwidth,angle=270,clip]{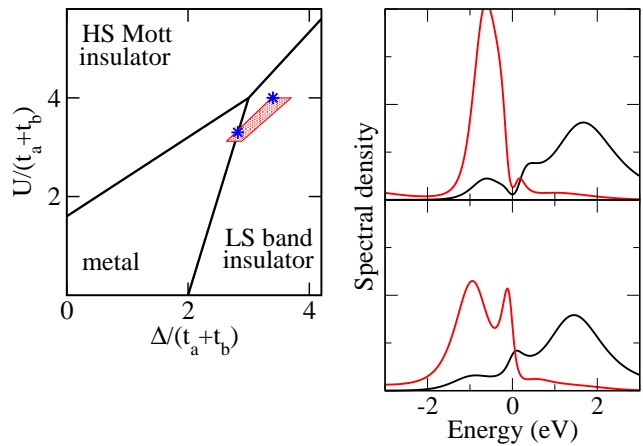}
\caption{ \label{fig:diag} (color online) Left: the conceptual phase diagram of the two-band
Hubbard model for $U=4J$. The shaded area marks the parameter range visited while varying the
band asymmetry $\zeta$ and crystal field $\Delta$. Right: 1P spectral densities
obtained at the points marked by stars (upper panel corresponds
to the upper star) at temperatures just above the leading $T_c$.}
\end{figure}

The connection to the bosonic physics arises in the strong-coupling
limit. As was shown by Batista~\cite{batista02}, the extended Falicov-Kimball model
at half filling maps onto spinless hard-core bosons with nn repulsion, a problem
much studied in the context of solid, superfluid and possibly
a supersolid phase~\cite{schmid02,bartouni00}.
We show that in the strong-coupling limit of the two-band Hubbard model with Hund's coupling
the mapping generalizes to the spinful hard-core bosons
with some additional nn terms, a much less studied problem
~\cite{kuklov04,boninsegni08} with a rich phase diagram.

The paper is structured as follows. In Section \ref{sec:comp} we
state the problem and describe the computational method. In Section \ref{sec:num}
we summarize our numerical results. In Section \ref{sec:disc}
we derive the strong- and weak-coupling limits of the studied model in order
to elucidate the nature of the instabilities reported in Section \ref{sec:num}.
We briefly discuss the classical limit, which provides the simple understanding
of the HS-LS phase, and then focus on various aspects of the excitonic 
phase. In Section \ref{sec:conc} we summarize our main findings.
\begin{figure}
  \begin{center}
    \includegraphics[width=0.55\columnwidth,angle=270,clip]{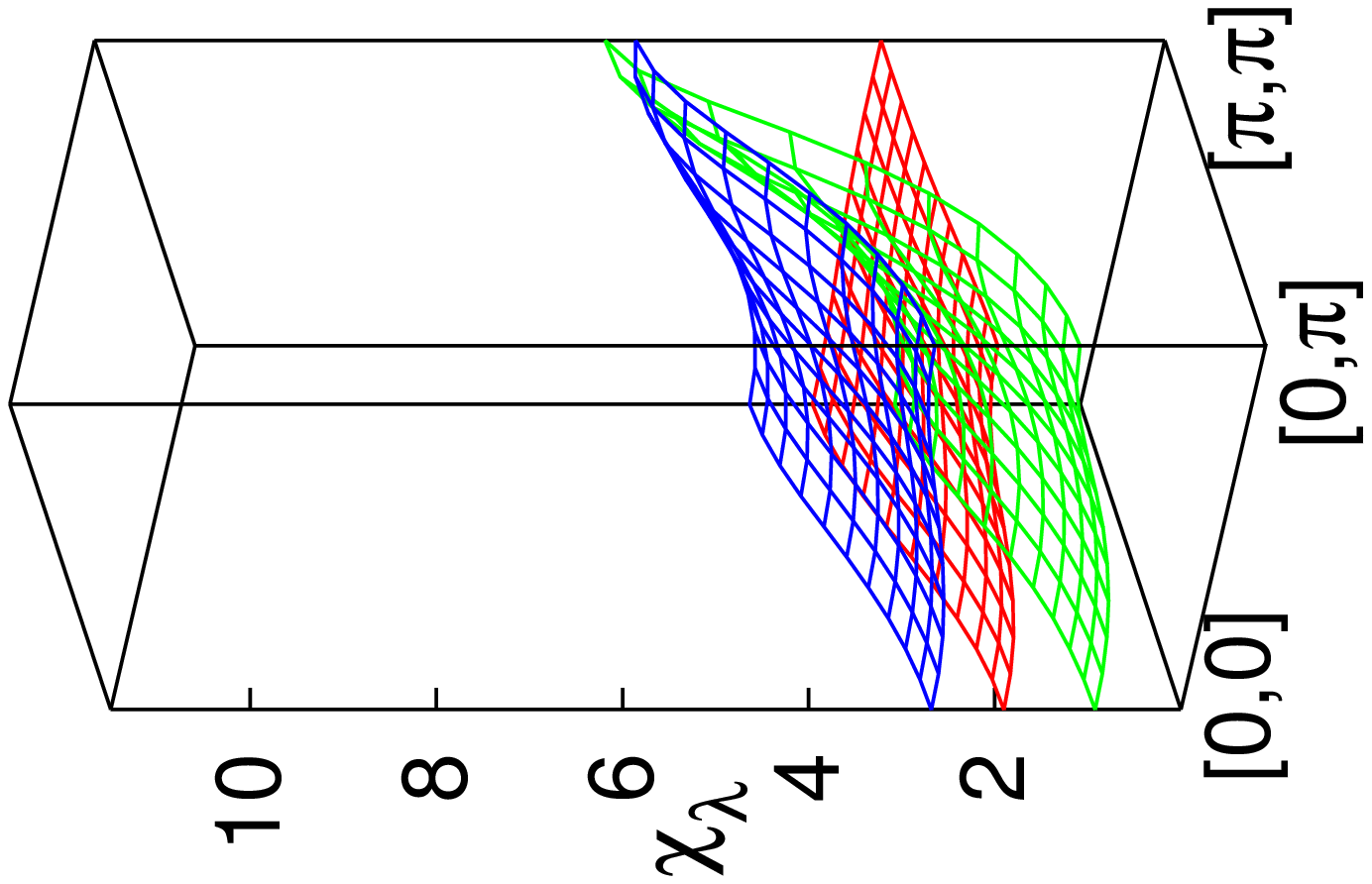}
    \includegraphics[width=0.55\columnwidth,angle=270,clip]{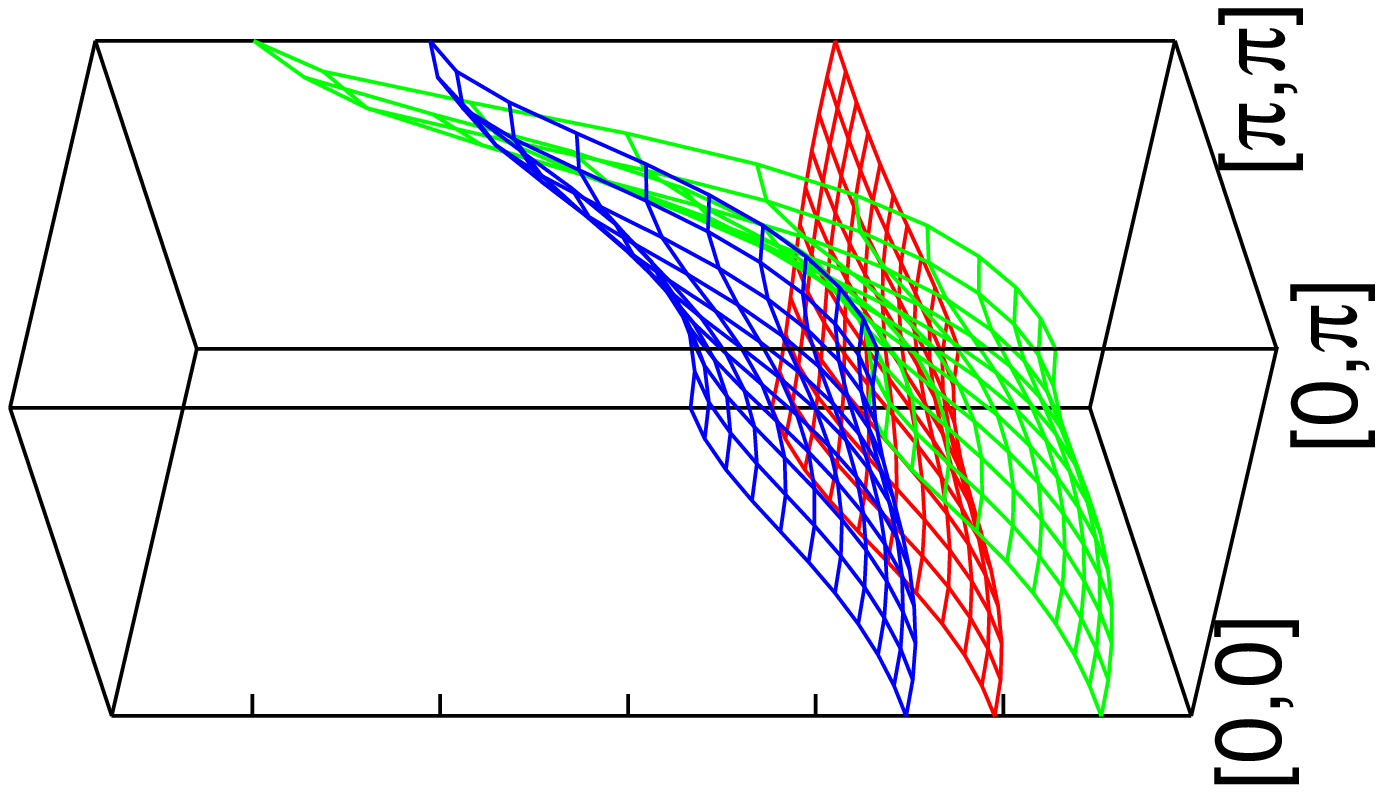}
    \includegraphics[width=0.55\columnwidth,angle=270,clip]{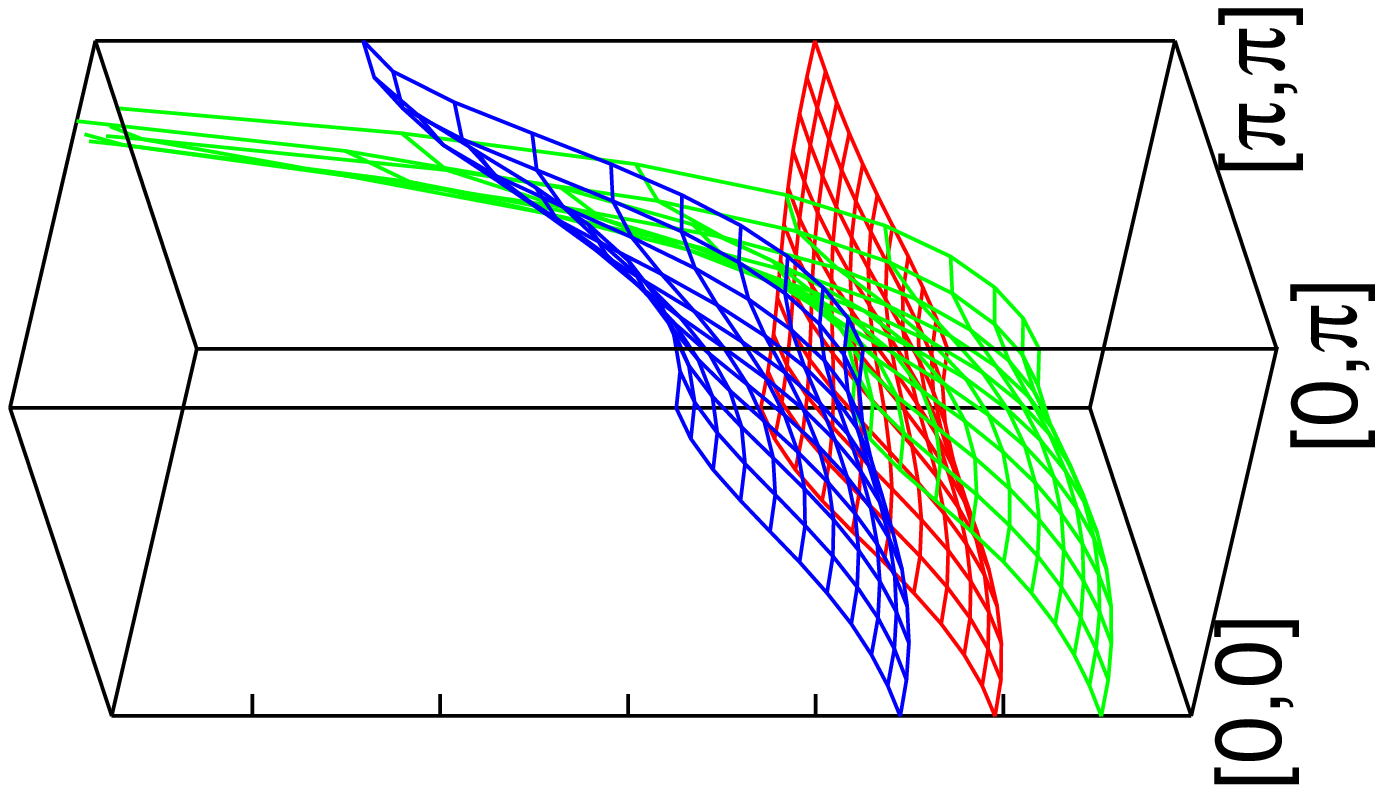}
  \end{center}
\caption{ \label{fig:susc1} (color online) The typical q-dependence of the leading eigenvalues
of the susceptibility matrix: spin longitudinal (red), OD (green) and OO (blue) in a system
with a large band asymmetry $\zeta=0.22$, $\Delta=3.40$ at temperatures 773~K, 644~K and 580~K
(left to right).}
\end{figure}

\section{\label{sec:comp}Computational procedure}
We consider the two-band Hubbard mode with nearest-neighbor (nn) hopping on a bipartite (square) lattice
with the kinetic
$H_{\text{t}}$ and the interaction $H_{\text{int}}=H^{\text{dd}}_{\text{int}}+H'_{\text{int}}$ terms given by
\begin{equation}
\label{eq:hubbard}
\begin{split}
&H_{\text{t}}=\frac{\Delta}{2}\sum_{i,\sigma}\bigl(n^a_{i\sigma}-n^b_{i\sigma}\bigr)+
  \sum_{i,j,\sigma}\bigl(t_{a} a_{i\sigma}^{\dagger}a^{\phantom\dagger}_{j\sigma}+
t_{b} b_{i\sigma}^{\dagger}b^{\phantom\dagger}_{j\sigma}\bigr) \\
&\qquad+\sum_{\langle ij\rangle,\sigma}\bigl(V_1a_{i\sigma}^{\dagger}b^{\phantom\dagger}_{j\sigma}+
V_2b_{i\sigma}^{\dagger}a^{\phantom\dagger}_{j\sigma}+c.c.\bigr) \\
&H^{\text{dd}}_{\text{int}}=U\sum_i \bigl(n^a_{i\uparrow}n^a_{i\downarrow}+n^b_{i\uparrow}n^b_{i\downarrow}\bigr)+
  (U-2J)\sum_{i,\sigma} n^a_{i\sigma}n^b_{i-\sigma}\\
&\qquad+(U-3J)\sum_{i\sigma} n^a_{i\sigma}n^b_{i\sigma} \\
&H'_{\text{int}}= J \sum_{i\sigma} a_{i\sigma}^{\dagger}b_{i-\sigma}^{\dagger}
a_{i-\sigma}^{\phantom\dagger}b_{i\sigma}^{\phantom\dagger} 
+J'\sum_{i} \bigl(a_{i\uparrow}^{\dagger}a_{i\downarrow}^{\dagger}b_{i\downarrow}^{\phantom\dagger}
b_{i\uparrow}^{\phantom\dagger}+c.c.\bigr).
\end{split}
\end{equation}
Here $a_{i\sigma}^{\dagger}$, $b_{i\sigma}^{\dagger}$
are the creation operators of fermions with spin $\sigma=\uparrow,\downarrow$
and $n^c_{i\sigma}=c_{i\sigma}^{\dagger}c^{\phantom\dagger}_{i\sigma}$.
Symbol $\sum_{i,j}$ implies summation over ordered nn pairs, while
$\sum_{\langle ij\rangle}$ implies summation over nn bonds (pairs without order).
The model is studied at half filling, two electrons per site on average. The crystal field
$\Delta$ and the Hund's exchange $J$ are chosen so that the system is in the vicinity
of the LS-HS transition.

The numerical calculations were performed in the dynamical mean-field approximation~\cite{dmft,metzner89}
with the density-density interaction $H^{\text{dd}}_{\text{int}}$ only.
The effect of adding $H'_{\text{int}}$ is considered in Section \ref{sec:disc}.
We use the hybridization expansion continuous time quantum
Monte Carlo (CT-HYB)~\cite{werner06,ctqmc} to solve the auxiliary impurity
problem and obtain the local one-particle (1P) and two-particle (2P) propagators.
For selected parameters we have benchmarked the CT-HYB results against those
obtained with the Hirsch-Fye implementation of the present procedure~\cite{kunes11}.

In order to study phase transitions, we search numerically for divergent static 
particle-hole susceptibilities in the disordered high temperature phase.
The lattice susceptibility $\chi_{\alpha\beta,\gamma\delta}(T,\mathbf{q})$ 
is a $\mathbf{q}$-dependent matrix function indexed by pairs of spin-orbital indices. 
It is calculated from the Bethe-Salpeter equation as a function
of the full 1P propagator and the 2P-irreducible vertex.
The crucial DMFT simplification consists in the fact that the 2P irreducible vertex 
is $\mathbf{k}$-independent and equals the impurity 2P irreducible vertex~\cite{dmft}. 
Therefore the momentum dependence of $\chi(T,\mathbf{q})$ comes entirely from the 1P propagator.

We calculate $\chi(T,\mathbf{q})$ on  dense $\mathbf{q}$-mesh in the Brillouin
zone, diagonalize for every $\mathbf{q}$, and identify the largest eigenvalues with
the corresponding eigenvectors.
The transition temperature is obtained from the zero crossing
$\chi^{-1}_{\lambda}(T_c)=0$ of the inverse of the largest eigenvalue
$\chi^{-1}_{\lambda}(T,\mathbf{q})=0$.
The advantage of this approach is that no prior assumptions about 
the symmetry of the ordered phase is needed.

\begin{figure} 
  \begin{center}
    \includegraphics[width=0.55\columnwidth,angle=270,clip]{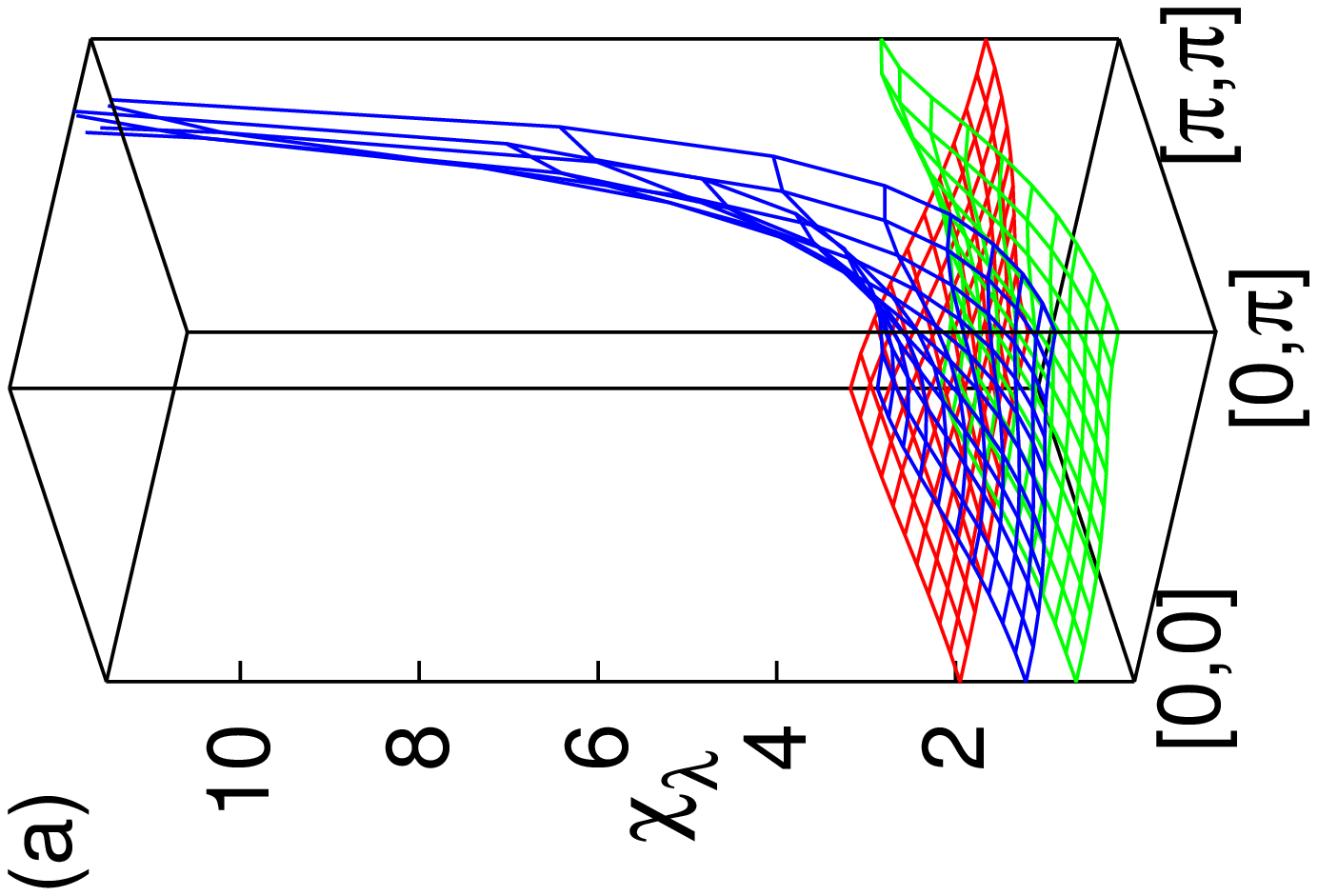}
    \includegraphics[width=0.55\columnwidth,angle=270,clip]{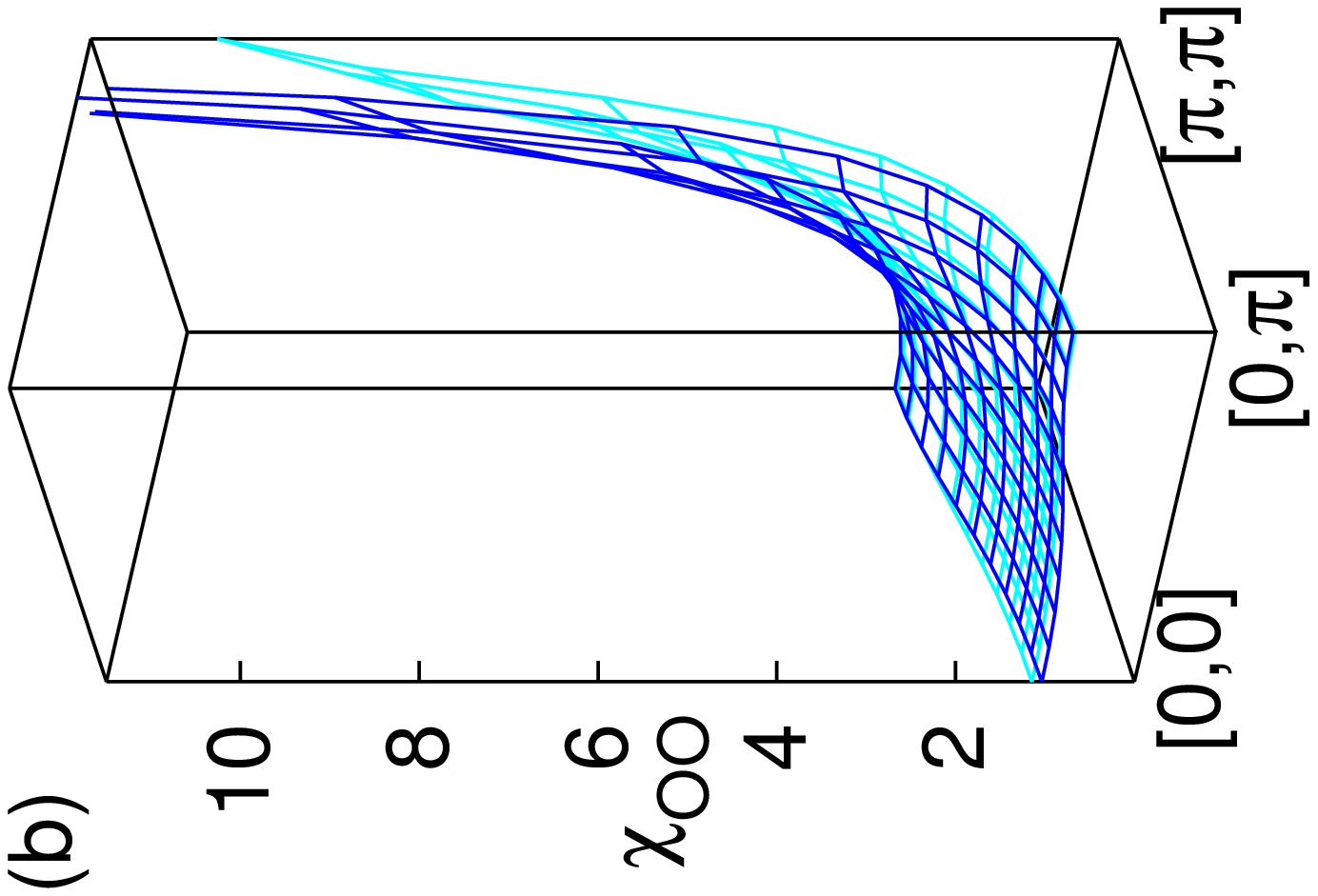}
  \end{center}
\caption{ \label{fig:susc2} (color online)
Left: Leading eigenvalues for equal bandwidths ($\zeta=1$) and $\Delta=3.40$~eV at 1160~K. The blue OO
mode diverges faster than the green OD mode.
Right: Splitting of the OO mode from (b) due to added cross-hopping $V_{1,2}=0.1~eV$.
The leading mode (two-fold degenerate) has the form $a^{\dagger}_{\sigma}b_{-\sigma}+
b^{\dagger}_{\sigma}a_{-\sigma}$ with $\sigma=\uparrow$, $\downarrow$.
}
\end{figure}
\section{\label{sec:num}Numerical results}
In this section we present the DMFT results obtained for
the Hamiltonian $H_t+H^{\text{dd}}_{\text{int}}$.
Following Ref.~\onlinecite{kunes11}, we set $U$=4, $J$=1 and 
use eV as energy units to allow for a straightforward comparison.
The basic phase diagram of model (\ref{eq:hubbard}) at half filling was 
computed by Werner and Millis~\cite{werner07} and its cartoon version
is presented in Fig.~\ref{fig:diag}. 
We are interested in a small region close to the boundary between
HS Mott insulator and LS band insulator, which fixes the $\Delta$
of interest to $3J$ approximately.
Our main variable parameter will be the asymmetry between $a$ and $b$ derived
band characterized by $\zeta=\tfrac{2t_at_b}{t_a^2+t_b^2}$.
For reason that becomes apparent in the discussion 
of the strong coupling limit, we choose to vary $\zeta$ while keeping 
the sum $t_a^2+t_b^2$ fixed.
Consequently, the point representing our system moves slightly, covering
the red region of Fig.~\ref{fig:diag} when going between symmetric bands, $\zeta=1$, and the 
flat-band limit, $\zeta=0$.

\begin{figure}
    \includegraphics[height=0.48\columnwidth,angle=270,clip]{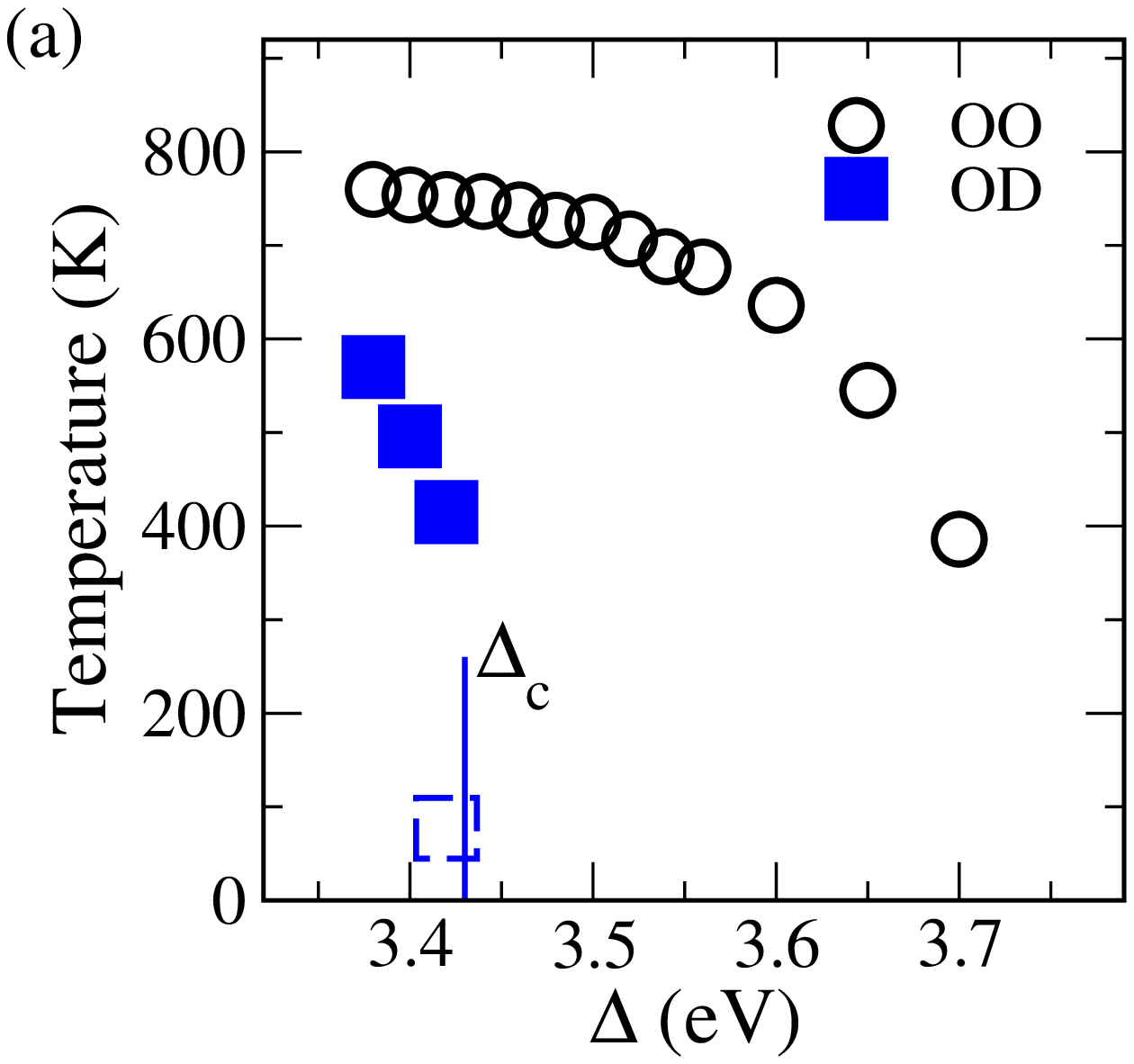}
    \includegraphics[height=0.48\columnwidth,angle=270,clip]{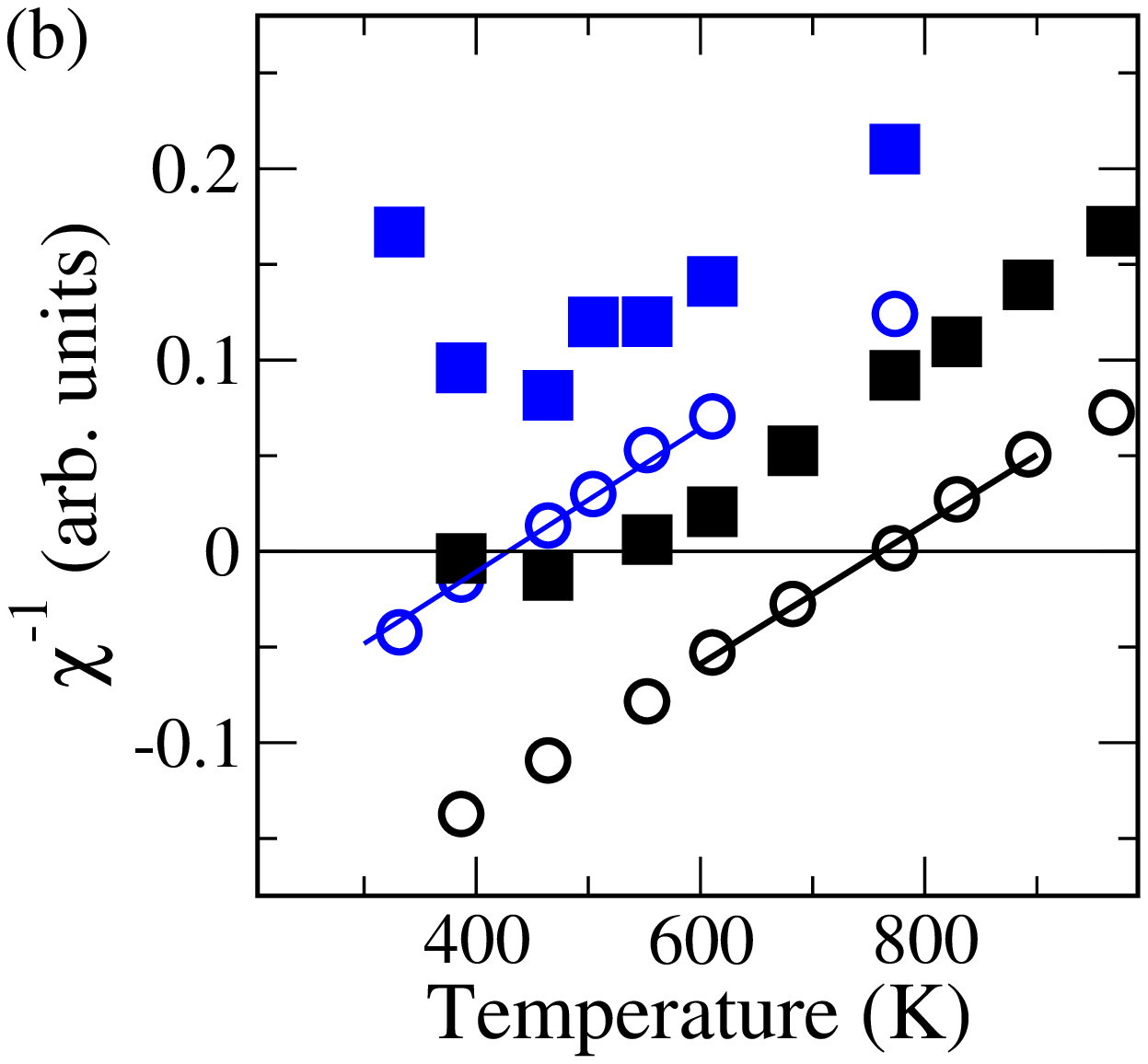}
\caption{\label{fig:delta}(color online)
(a) Representative dependencies of the instability
temperatures on the crystal field $\Delta$: $T_{\text{OD}}$ (squares) for $\zeta=0.28$ and
$T_{\text{OO}}$ (circles) for $\zeta=0.55$. The open square marks the position of the
reentrant transition taken from Ref.~\onlinecite{kunes11}. The blue line marks the
estimated position of $\Delta_c$.
(b) The $T$-dependence of the inverse eigenvalues $\chi^{-1}_{\text{OO}}$ (circles) and
$\chi^{-1}_{\text{OD}}$ (squares) of the susceptibility at selected values of $\Delta$.
The parameters  $\zeta=0.28$, $\Delta=3.44$~eV (blue) correspond to $\Delta\gsim\Delta_c$
where the OD instability already disappeared. For $\zeta=0.55$, $\Delta=3.38$~eV (black)
the OD instability exists only in a finite interval of temperatures. In both cases
the OO is the leading instability, which is physically realized.}
\end{figure}
First, we discuss the eigenmodes of $\chi(\mathbf{q})$ for
$t_a=0.45$~eV, $t_b=0.05$~eV ($\zeta=0.22$), $V_{1,2}=0$, and $\Delta=3.40$~eV, 
the parameters of Ref.~\onlinecite{kunes11}.
The full $16\times16$ matrix of $\chi(\mathbf{q})$
can be, in a standard way using the spin-conservation law, block-diagonalized to
$\uparrow\uparrow-\downarrow\downarrow$, $\uparrow\uparrow+\downarrow\downarrow$,
$\uparrow\downarrow$ and $\downarrow\uparrow$ blocks (channels), each having
$4\times4$ orbital structure.
We find three distinct branches of $\chi_{\lambda}(\mathbf{q})$
with  sizable magnitude. These correspond to 
i) the spin longitudinal mode $\sum_{\sigma}\sigma(n^a_{\sigma}+n^b_{\sigma})$
in the $\uparrow\uparrow-\downarrow\downarrow$ channel,
ii) the orbital diagonal (OD) mode $\sum_{\sigma}(n^a_{\sigma}-n^b_{\sigma})$
in the $\uparrow\uparrow+\downarrow\downarrow$ channel, and
iii) four degenerate orbital off-diagonal (OO) modes
$a^{\dagger}_{\uparrow}b^{\phantom\dagger}_{\downarrow}$,
$b^{\dagger}_{\uparrow}a^{\phantom\dagger}_{\downarrow}$,
$a^{\dagger}_{\downarrow}b^{\phantom\dagger}_{\uparrow}$,
$b^{\dagger}_{\downarrow}a^{\phantom\dagger}_{\uparrow}$ in the
$\uparrow\downarrow$ and $\downarrow\uparrow$ channels.
In Fig.~\ref{fig:susc1}, the $\mathbf{q}$ dependence of the corresponding eigenvalues 
in the 2D Brillouin zone is plotted for several temperatures. 
Similar plot for symmetric bands, $\zeta=1$, is shown in Fig.~\ref{fig:susc2}.

The leading instability for $\zeta=0.22$ is identified in the OD mode
at $(\pi,\pi)$. The corresponding transition temperature agrees well
with the onset of the HS-LS checker-board order found in Ref.~\onlinecite{kunes11}. 
Increasing the crystal field $\Delta$ rapidly suppresses the transition temperature
$T_{OD}$, see Fig.~\ref{fig:delta}a, and the OD instability eventually disappears
above some $\Delta_c$. For $\Delta\lsim\Delta_c$ the OD instability disappears
at low temperatures as shown in Fig.~\ref{fig:delta}b, leading to a reentrant transition.
For $\Delta\gsim\Delta_c$, the proximity of the ordered phase at 
an intermediate temperature gives rise to a peak in the susceptibility, Fig.~\ref{fig:delta}b.
These results provide the same picture as the calculations of Ref.~\onlinecite{kunes11}
performed in the ordered HS-LS phase. However, in addition to that, one can see
that the OO susceptibility also exhibits a substantial increase at $(\pi,\pi)$
with decreasing temperature.

Next, we vary the band asymmetry $\zeta$ while keeping the cross-hopping $V_{1,2}=0$.
For more symmetric bands a different result is obtained, as shown in Fig.~\ref{fig:susc2},
where the dominant $\chi_{\lambda}(\mathbf{q})$ are plotted for $\zeta=1$. 
In this case, the OO mode at $(\pi,\pi)$ is the leading instability. This implies formation of 
an ordered state with spontaneous local off-diagonal hybridization characterized by non-zero value
of $\langle a^{\dagger}_{i,\sigma}b^{\phantom\dagger}_{i,-\sigma}\rangle$ and anti-ferro periodicity.

\begin{figure}
    \includegraphics[height=0.7\columnwidth,angle=270,clip]{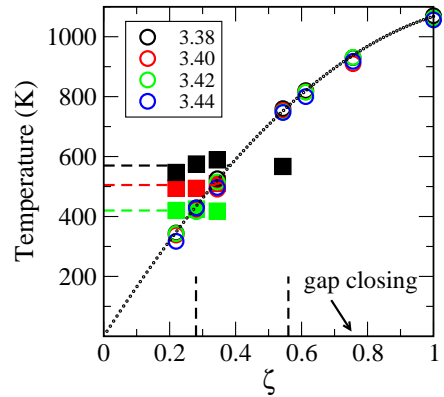}
\caption{\label{fig:phase}(color online) Instability of the normal phase
as a function of band asymmetry $\zeta$ for various CF parameters $\Delta$. Open circles
denote the divergence $T_{\text{OO}}$ of the OO mode, filled squares mark
the divergence $T_{\text{OD}}$ of the OD mode. The lines are guides to the eye.
The dashed vertical lines mark the $\zeta$'s for which the $\Delta$ dependences
of $T_{\text{OO}}$ and $T_{\text{OD}}$ are shown in Fig.~\ref{fig:delta}a.
}
\end{figure}
In Fig.~\ref{fig:phase}, we show the calculated instability lines in the $\zeta$-$T$ plane 
for several values of $\Delta$. The actual calculations were performed for $t_b\leq t_a$, but 
the results hold also for $t_a\leq t_b$, since on a bipartite lattice at half-filling the latter
can be mapped on the former by exchange of $a$ and $b$ followed by the particle-hole transformation
and the sign reversal of $a$ and $b$ operators on one sublattice. 
Several observations can be made. For the studied parameters there are two possible instabilities
corresponding to the OO and OD modes. The OO mode, favored by more symmetric bands, is the leading
instability over a broad range of band asymmetries. The OO instability is suppressed when one of the 
bands becomes narrow, in which case the instability line $T_{OO}(\zeta)$ extrapolates linearly to zero.
The OD mode is the leading instability only for strongly asymmetric bands.
For constant $t_a^2+t_b^2$, the $T_{OD}(\zeta)$ is insensitive to $\zeta$ within the 
accuracy of our calculation. For all $\zeta$, the $T_{OO}$ is less sensitive to 
the crystal field $\Delta$ than $T_{OD}$.

The OO instability shows little sensitivity to the presence of a charge gap in the disordered state as
there is no apparent change in the behavior of $T_{OO}(\zeta)$ when the gap disappears.
In Fig.~\ref{fig:phase}, we mark closing of the charge gap above the LS state. 
The actual 1P spectral functions at temperatures just above $T_c$ close to both ends of the $\zeta$-range 
are shown in Fig.~\ref{fig:diag}.

The results obtained for positive $t_a$ and $t_b$ can be readily extended to 
an arbitrary combination of $\pm t_a$, $\pm t_b$ by the transformation 
$c_i\rightarrow (-1)^{i}c_i$ ($c=a$ and/or $b$). This is because for
$V_{1,2}=0$ the orbital diagonal and orbital off-diagonal modes do not mix
even within the same channel. The OD susceptibility $\chi_{OD}(\mathbf{q})$ is then insensitive 
to the signs of $t_a$ and $t_b$, i.e. the OD divergence
always takes place at $(\pi,\pi)$. The OO susceptibility $\chi_{OO}(\mathbf{q})$
is shifted by $(\pi,\pi)$ if $t_at_b<0$, i.e. the OO divergence is at the zone center
in this case.

For small non-zero cross hopping $V_{1,2}$ the location of the divergent modes are
still determined by the signs of $t_a$ and $t_b$. The main effect of such a finite
$V_{1,2}$ is a partial lifting of the degeneracy of $\chi_{OO}(\mathbf{q})$, as
shown in Fig.~\ref{fig:susc2} for $V_{1,2}=0.1$~eV.
The $a^{\dagger}_{\sigma}b^{\phantom\dagger}_{-\sigma}$
and $b^{\dagger}_{\sigma}a^{\phantom\dagger}_{-\sigma}$ modes form
symmetric and anti-symmetric combinations which follow distinct $\mathbf{q}$ dependences.
The degeneracy of $\uparrow\downarrow$ and $\downarrow\uparrow$ channels 
is not affected by the spin preserving hopping. 

\section{\label{sec:disc}Discussion}
Before discussing various limits of the studied model, we point out
formal equivalence between the excitonic condensation and superconductivity.
This can be seen by exchanging the notion of particle and hole for one of the
fermionic species, e.g. $b_i^{\phantom\dagger}\rightarrow b_i^{\dagger}$, which turns 
$a$-$b$ repulsion into attraction. This equivalence obviously breaks down
when electromagnetic response is concerned since the excitons carry no charge.
Nevertheless, it is useful to consider the analogy to superconductivity, which
is more familiar to most physicists. The excitonic order parameter in our study
is local, i.e. has no $\mathbf{k}$-dependence, which is analogous
to $s$-wave superconductivity. 
An $\langle ab \rangle$ order parameter, composed of different orbitals,
is unusual for a superconductor, due to the weakness of the electron-electron attraction,
but can be easily realized in an excitonic condensate, as
the electron-hole attraction is strong.
Consisting of two distinct orbitals, the spin part $\langle ab \rangle$ order parameter
is not restricted by Pauli principle and can be both singlet or triplet.
It is the $J>0$ Hund's coupling which selects the triplet parameter in the studied
model. Like for superconductivity, one may consider the strong-coupling (BEC)
and the weak-coupling (BCS) limits.

\subsection{Strong-coupling limit}
The strong-coupling limit is characterized by
the LS and HS states being separated from the remaining atomic states
by energy $E_i-E_{\text{HS}/\text{LS}}\gg |t_a|, |t_b|, |V_{1,2}|$. In this 
case an effective model without charge fluctuations can be formulated 
using the Schrieffer-Wolff transformation \cite{schrieffer}, which
provides a simplified picture of the low-energy physics.
The resulting effective Hamiltonian with hopping treated to the second
order is derived in Appendix A. In the following, we discuss some of 
its aspects.

\subsubsection{Density-density interaction ($H'_{\text{int}}=0$)}
First, we consider model (\ref{eq:hubbard}) with the density-density interaction only
for which the DMFT calculations, reported in preceding section, were performed.
The effective Hamiltonian then has the form 
\begin{equation}
\label{eq:boson1}
\begin{split}
H^{\text{dd}}_{\text{eff}}&=\sum_{i} \mu n_i+K_{\perp}\sum_{ij,s}d^{\dagger}_{i,s}d^{\phantom\dagger}_{j,s}
               +\sum_{\langle ij \rangle}\bigl(K_{\parallel}n_in_j
               + K_0S^z_i S^z_j\bigr) \\
&+K_1\sum_{\langle ij \rangle,s} \left(
d^{\dagger}_{i,s}d^{\dagger}_{j,-s}+d^{\phantom\dagger}_{i,s}
d^{\phantom\dagger}_{j,-s}\right).
\end{split}
\end{equation}
describing two flavors $s=\pm1$ of bosons with the hard-core constraint 
$n_i=\sum_{s}d^{\dagger}_{i,s}d^{\phantom\dagger}_{i,s}\le 1$,
corresponding to HS states created by
$d^{\dagger}_{1}=a^{\dagger}_{\uparrow} b^{\phantom\dagger}_{\downarrow}$ and
$d^{\dagger}_{-1}=a^{\dagger}_{\downarrow}b^{\phantom\dagger}_{\uparrow}$ out 
of the LS vacuum.
Neglecting the cross-hopping contribution the coupling constants have a simple
form $\mu=\Delta-3J-Z\tfrac{t_a^2+t_b^2}{U-2J}$, $K_{\perp}=\frac{2t_at_b}{U-2J}$,
$K_{\parallel}=(t_a^2+t_b^2)\tfrac{U+4J}{(U-2J)(U+J)}$, and
$K_0=\tfrac{t_a^2+t_b^2}{U+J}$, where $Z=4$ is the number of nearest neighbors.
The last term appears only for finite cross hopping and
has the form $K_1=-2V_1V_2\tfrac{U-2J}{(U+J-\Delta)(U-5J+\Delta)}$.

\subsubsection{Classical limit ($\zeta=0$)}
The behavior of model (\ref{eq:hubbard}) as revealed by the DMFT calculations strongly depends 
on the band asymmetry $\zeta$.
The OD instability was found only for rather asymmetric bands $t_at_b\ll t_a^2+t_b^2$, 
which leads to $K_{\perp}\ll K_{\parallel}$ in (\ref{eq:boson1}). 
In the limit $t_at_b=0$ the hopping $K_{\perp}$ disappears, and the effective
model (\ref{eq:boson1}) reduces to the classical
Blume-Emmery-Griffiths (BEG) model~\cite{beg}.
Assigning $s_i=0$ to $|\text{LS}\rangle$ and $s_i=\pm 1$ 
to $d^{\dagger}_{\pm 1}|\text{LS}\rangle$ one arrives at its usual form
\begin{equation}
\label{eq:beg}
H_{\text{BEG}}=\mu\sum_{i}s_i^2+\sum_{\langle ij \rangle}\bigl(K_{\parallel}s_i^2 s_j^2 +
  K_0s_i s_j\bigr).
\end{equation} 
With our choice of the parameters $U$, $J$, $t_a^2+t_b^2$ and $\Delta$, we have $\mu\approx0$
($\mu=0$ corresponds to $\Delta=3.41$) and  $K_{\parallel}/K_0=4$.
According to Ref.~\onlinecite{hoston91}, 
for $K_{\parallel}/K_0=4$ and $\mu$ between $\mu_{\text{min}}<0<\mu_{\text{max}}$
the BEG model exhibits a solid (S) order, characterized by a checker-board arrangement of HS and LS sites.
This is equivalent to a staggered density $\langle n_i \rangle$ in the language of the bosonic model (\ref{eq:boson1}).
For $\mu<0$ the order exists down to the zero temperature, for $\mu>0$ the order disappears at finite $T$.
The solid order as well as the reentrant transition was found also in previous DMFT simulations~\cite{kunes11}
of 2BHM with asymmetric bands. Proximity to the BEG limit thus provides a simple explanation
of the OD instability in the strong coupling and asymmetric bands region of model (\ref{eq:hubbard}). 
The analysis of the BEG model~\cite{hoston91} suggests that for $\mu\approx\mu_{\text{min}}$
competition between the anti-ferromagnetic and the solid phase gives rise to a rather complicated
phase diagram. This parameter range is, however, beyond the scope of this work.

\subsubsection{Superfluid phase}
For general $\zeta$, the hopping $K_{\perp}$ cannot be neglected.
Much studied in the context of cold atoms, the spinless version of (\ref{eq:boson1})
is known to host a superfluid (SF) phase in addition to the solid (S) phase discussed above.
Existence of a supersolid order at the boundary between S and SF phases is a subject of 
intense research on the model generalizations~\cite{mila08}.
The spinless model (\ref{eq:boson1}) can also be derived 
as the strong-coupling limit of the extended Falicov-Kimball model~\cite{batista02}.

The SF phase is characterized by a finite value of $\langle d_{i,s} \rangle$,
which corresponds to spontaneous appearance of an off-diagonal 
expectation value $\langle a^{\dagger}_{i,\sigma}b^{}_{i,-\sigma}\rangle$ in 2BHM,
and thus can be identified with the observed OO instability.
Without cross-hopping, $V_{1,2}=0$, the SF phase of (\ref{eq:boson1}) is similar to 
the spinless case in the sense that it consists of two copies of the latter coupled only by
amplitude fluctuations. Inclusion of the cross-hopping has a very different effect
on the spinless and spinful models.
In the spinless case~\cite{batista02}, the cross-hopping must have the same 
form as the $d$ operator and thus non-zero $V$ introduces a source 
term $V^*d+Vd^{\dagger}$ to the Hamiltonian, removing the distinction between the 
normal and SF phases. 
In the spinful case (\ref{eq:boson1}), however, the spin-preserving cross hopping has 
a different spin symmetry than the $d_s$ operators and therefore non-zero $V$ introduces the $K_1$ term instead. 
Finite $K_1$ locks together the phases of $\langle d_{i,1}\rangle$ 
and $\langle d_{i,-1}\rangle$. This is reflected in the partial lifting of the degeneracy of the OO mode.
The distinction between the normal and SF phases is thus preserved irrespective of the
cross hopping.

\subsubsection{SU(2) symmetric interaction}
Next, we discuss the effect of the spin-flip and pair-hopping terms in $H'_{\text{int}}$,
which were not included in the DMFT simulation.
The spin-flip term renders model (\ref{eq:hubbard}) $SU(2)$ symmetric and a third boson
$d^{\dagger}_{0}= \tfrac{a^{\dagger}_{\uparrow}b^{\phantom\dagger}_{\uparrow}-
a^{\dagger}_{\downarrow}b^{\phantom\dagger}_{\downarrow}}{\sqrt{2}}$ 
appears in the effective model
\begin{equation}
\label{eq:boson2}
\begin{split}
H_{\text{eff}}=&\sum_{i} \mu n_{i}+K_{\perp}\sum_{ij}\mathbf{d}^{\dagger}_{i}\mathbf{d}^{\phantom\dagger}_{j}\\
               +&\sum_{\langle ij \rangle}\bigl(K_{\parallel}n_in_j
               + K_0\operatorname{\mathbf{S}_i}\cdot \operatorname{\mathbf{S}_j}\bigr)\\
               -&K_1\sum_{\langle ij \rangle} \left(\mathbf{d}^{\dagger}_{i}\cdot\mathbf{d}^{\dagger}_{j}+\mathbf{d}^{\phantom\dagger}_{i}\cdot\mathbf{d}^{\phantom\dagger}_{j}\right)\\
               +& K_2 \sum_{i,j}(\mathbf{d}^{\phantom\dagger}_{i}+\mathbf{d}^{\dagger}_{i})
                \cdot\operatorname{\mathbf{S}}_j.
\end{split}
\end{equation}
Here, $\operatorname{(\mathbf{S}_i)_{\alpha}}=\sum_{ss'}d^{\dagger}_{i,s}S^{\alpha}_{ss'}d^{\phantom\dagger}_{i,s'}$
and $n_i=\sum_{s}d^{\dagger}_{i,s}d^{\phantom\dagger}_{i,s}$,
where $s=0,\pm1$ and $S^{\alpha}_{ss'}$ are spin S=1 operators.
The $d$ operators are arranged in a vector
$\mathbf{d}=\bigl(\tfrac{1}{\sqrt{2}}(d_{-1}-d_1),\tfrac{1}{i\sqrt{2}}(d_{-1}+d_1),d_0\bigr)$.
As before, the hard-core constraint $n_i\leq$ is assumed. 
We are not aware of any specific studies of the $S=1$ model (\ref{eq:boson2}).
On a mean field level one can repeat the arguments used for the
density-density interaction which lead to the expectation of solid order
for $K_{\perp}\ll K_{\parallel}$. The SF order parameter 
generalizes to a 3-component vector the phase of which 
is again determined by the $K_1$ term. The $K_2$ term is new 
and does not have an analogy in the density-density case. 

\subsubsection{Coupling constants}
The full expressions for the coupling constants are given in Appendix A.
Here, we consider their signs as functions the hopping parameters
$t_{a,b}$ and $V_{1,2}$ and implications for the broken symmetry phases.

Varying the chemical potential $\mu\approx\Delta-3J$,
we can tune between two `trivial' limits: the vacuum state $\langle n_i\rangle\approx 0$ for large
$\Delta$ corresponding to the LS grounds tate of (\ref{eq:hubbard}) and
$\langle n_i\rangle\approx 1$ for small $\Delta$, which corresponds to anti-ferromagnetic S=1 Heisenberg model.  
Our DMFT calculations fall into the intermediate $\Delta$ regime with non-integer $\langle n_i\rangle$.

The fact that $K_{\parallel}$ is always positive, being proportional to $t_a^2+t_b^2,V_1^2+V_2^2$, 
implies that, irrespective of the signs of the hoppings, the OD instability 
leads always to an anti-ferro (AF) order.
Similarly, $K_0\sim t_a^2+t_b^2,V_1^2+V_2^2$ implies that there is always AF magnetic
interaction between the nearest neighbors.
The sign of $K_{\perp}\sim t_at_b$ depends on the relative sign of $t_a$ and $t_b$.
The cross-hopping contribution to $K_{\perp}$ is proportional to $V_1V_2J'$ and thus may interfere both 
constructively or destructively with the $t_at_b$ term.
$K_{\perp}>0$ favors AF SF order while $K_{\perp}<0$ favors ferro (F) SF order
on a given bond. Therefore the OO divergence can be moved from $(\pi,\pi)$
to $(0,0)$ simply by changing the sign of $t_a$ or $t_b$. 

Non-zero $K_1$ fixes the phase of $\langle \mathbf d\rangle$ in the SF phase.
Depending on the sign of $K_1K_{\perp}$ it selects $\langle \mathbf d\rangle$ to be real or imaginary. 
This corresponds to divergence of either the symmetric $a^{\dagger}b+ b^{\dagger}a$ 
or the anti-symmetric $a^{\dagger}b- b^{\dagger}a$ OO mode. The $K_1$ term appears
when the pair-hopping $J'\neq0$ or the cross-hopping $V_{1,2}\neq0$ is present.  
Inspection of the formulas in Appendix A shows that for $V_{1,2}=0$ 
the $K_1\sim -J't_at_b$ contribution always favors real $\langle \mathbf d\rangle$,
while for $V_{1,2}\neq0$ one can get either sign of $K_1K_{\perp}$.

Finally, $K_2\sim (V_1t_a+V_2t_b)$ appears only in the $SU(2)$ symmetric case
with the cross-hopping present.
In case of $\langle \mathbf d\rangle$ having a real component
this term acts as an effective Zeeman field and induces
spin polarization along $\langle \mathbf d\rangle$. 

\subsection{Weak-coupling limit}
In the weak coupling limit, we consider almost empty (full) $a$ ($b$) bands
with a small mutual overlap and search for the divergencies
of the static susceptibility using the random phase approximation.
The bare susceptibility, in this case, is dominated by the
diagonal elements $\chi^0_{ab,ab}$, corresponding to formation
of electron-hole pairs with different orbital characters.
The $\chi^0_{aa,aa}$ and $\chi^0_{bb,bb}$ elements, as
well as $\chi^0_{aa,ab}$ which may appear due to the cross-hopping,
are small and
we can restrict our considerations to the $2\times 2$ block of mixed orbital flavors.
Depending on the sign of $t_at_b$ the diagonal element $\chi^0_{ab,ab}$ 
is peaked either at $(0,0)$ or $(\pi,\pi)$ due to Fermi surface nesting. 
If $V_{1,2}\neq0$ an off-diagonal element $\chi^0_{ab,ba}$ appears.

We find divergent susceptibilities in the magnetic (triplet) channel
which have the form
\begin{equation}
\label{eq:rpa}
\chi^{S,A}_{\text{OO}}=\frac{\chi^0_{ab,ab}\pm\chi^0_{ab,ba}}{1-(U-2J\pm J')(\chi^0_{ab,ab}\pm\chi^0_{ab,ba})}
\end{equation}
and belong to a symmetric $a^{\dagger}b+b^{\dagger}a$ and an anti-symmetric
$a^{\dagger}b-b^{\dagger}a$ mode, respectively. 
Positive $J$ always favors $\chi^S_{\text{OO}}$ to be the leading
divergence. 
The cross-hopping $V_{1,2}$, which controls the sign of $\chi^0_{ab,ba}$, may select 
$\chi^S_{\text{OO}}$ as well as $\chi^A_{\text{OO}}$ to be the leading instability. 
For $J'=V_{1,2}=0$ the two modes are degenerate.
Without Hund's coupling~\cite{halperin68,balents00,zocher11,kaneko12} ($J=0$)
the singlet and triplet channels become degenerate.
In that case, non-zero cross-hopping $V_{1,2}$ may preclude the phase transition
in that the singlet excitonic pairing only enhances the existing off-diagonal
expectation values. With Hund's coupling the triplet order parameter always represents 
a true symmetry breaking as it has distinct symmetry for 
an arbitrary spin-preserving hopping. 

In the mean-field
picture, assuming an F order for simplicity, we get
\begin{equation}
\label{eq:mf}
H_{MF}=\begin{pmatrix} \varepsilon_a({\mathbf k})\sigma_0 & 
V({\mathbf k})\sigma_0 + \boldsymbol\sigma\cdot\boldsymbol\phi \\
V^*({\mathbf k})\sigma_0+\left(\boldsymbol\sigma\cdot\boldsymbol\phi\right)^* &
\varepsilon_b({\mathbf k})\sigma_0
\end{pmatrix},
\end{equation}
with $\sigma_{\alpha}$ being the Pauli matrices in the spin space.
Divergence of $\chi^S_{\text{OO}}$ implies $\boldsymbol\phi^*=\boldsymbol\phi$
while divergence of $\chi^A_{\text{OO}}$ implies $\boldsymbol\phi^*=-\boldsymbol\phi$ (for details see
Appendix B). 
Omitting the overall charge conservation, which is not broken at the transition,
the order parameter reduces the $SU(2)$ symmetry of (\ref{eq:hubbard})
into U(1) and thus behaves as a point on $S_2$ sphere.
If $J'=V=0$ Hamiltonian (\ref{eq:hubbard}) has additional $U(1)$ symmetry associated
with the relative phase of $a$ and $b$ states. Breaking this symmetry leads to a  
complex order parameter that lives in $S_1\times S_2$.

Expressions (\ref{eq:rpa}, \ref{eq:mf}) hold also in the case of density-density interaction with the 
provision that divergent $\chi^{S,A}_{\text{OO}}$ are found only in the $\uparrow\downarrow$
and $\downarrow\uparrow$ channels (not in $\uparrow\uparrow$-$\downarrow\downarrow$) and
$\phi_z\equiv0$ in (\ref{eq:mf}). The $SU(2)$ symmetry of Hamiltonian (\ref{eq:hubbard}) 
reduces to $U(1)$ in case of the density-density interaction.  
The order parameter for non-zero $V_{1,2}$ is a real or imaginary vector $(\phi_x,\phi_y)$
living in $S_1$. If $V_{1,2}=0$ the relative phases of all spin-orbital
flavors are independent leading to $[U(1)]^3$ symmetry, which is reduced
to $U(1)$ at the transition. The order parameter is then a complex vector $(\phi_x,\phi_y)$
living in $S_1\times S_1$.

\subsection{Physical meaning of the excitonic order parameter}
\begin{figure}
 \begin{center}
 \includegraphics[width=0.8\columnwidth,clip]{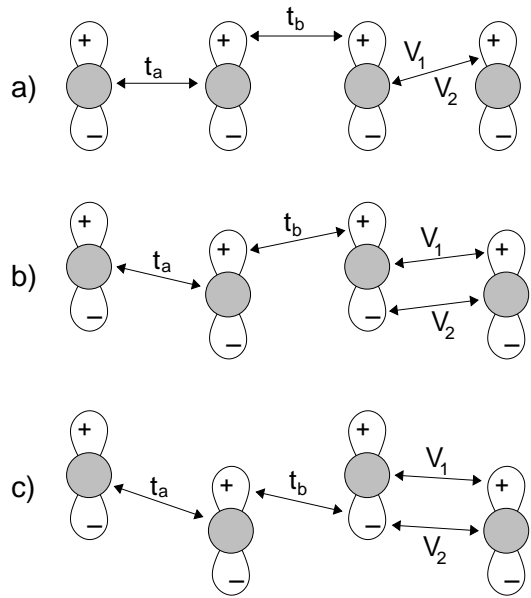}
 \end{center}
\caption{\label{fig:latt} 
An example of various combinations of the hoppings with 
orbitals of $s$ and $p_z$ symmetry.
$a)$ $t_{a,b}>0$, $V_{1,2}=0$,
$b)$ $t_{a,b}>0$, $V_1=-V_2$, and 
$c)$ $t_a>0$, $t_b<0$, $V_1=-V_2$.
}
\end{figure}

Finally, we discuss the physical meaning of the real, imaginary
or complex excitonic order parameter. In Fig.~\ref{fig:latt}
we present simple realizations of these phases using $s$ and $p_z$ orbitals:
a) $V_{1,2}=0$ with complex order parameter $\boldsymbol\phi$, b)
$t_at_bV_1V_2>0$ with real $\boldsymbol\phi$ and c) $t_at_bV_1V_2<0$
with imaginary $\boldsymbol\phi$. 

Let us start by considering real $\boldsymbol\phi=(0,0,\phi_z)$. The corresponding
operator $a^\dagger_{\uparrow}b^{\phantom\dagger}_{\uparrow}-
a^\dagger_{\downarrow}b^{\phantom\dagger}_{\downarrow}+b^\dagger_{\uparrow}a^{\phantom\dagger}_{\uparrow}
-b^\dagger_{\downarrow}a^{\phantom\dagger}_{\downarrow}$ describes the z-component
of magnetization (spin) density with the distribution given
by the product of $a$ and $b$ orbitals $\varphi_a(\mathbf{x})\varphi_b(\mathbf{x})$.
In present case, the product is a $p$ function, i.e. the leading multipole
of the distribution is a dipole and the above operator may be viewed as describing
an on-site magnetic quadrupole. The rotation of $\boldsymbol\phi$ corresponds
to changing the magnetization direction while keeping its distribution fixed, i.e.
cannot be viewed as a 3D rotation of the quadrupole as rigid object. 

The operator $a^\dagger_{\uparrow}b^{\phantom\dagger}_{\uparrow}-
a^\dagger_{\downarrow}b^{\phantom\dagger}_{\downarrow}-b^\dagger_{\uparrow}a^{\phantom\dagger}_{\uparrow}
+b^\dagger_{\downarrow}a^{\phantom\dagger}_{\downarrow}$ corresponding to
imaginary $\boldsymbol\phi=(0,0,\phi_z)$ describes an on-site pattern of a magnetization current.
Rotation of imaginary $\boldsymbol\phi$ corresponds to changing the
magnetization direction while keeping the current pattern fixed. 
Complex $\phi$ is difficult to visualize. In this case it is possible to continuously
rotate magnetic multipole into a local spin current without changing the
energy of the system. 

A model built on $d_{z^2}$ and $d_{x^2-y^2}$ orbitals may be more realistic with respect to 
real materials. Similar considerations would apply leading to a finite value
of magnetic octupole, in case of real, and more a complicated pattern of the on-site spin current,
in case of imaginary order parameter. While the direct experimental detection 
of the magnetic multipoles may be experimentally difficult, 
presumably, the most experimentally accessible would be the effect of excitonic
order on the transport properties at weak to moderate coupling.

\subsection{Further work}
Despite a narrow parameter range in the vicinity of the spin-state transition,
the present results reveal a rich phase diagram, nevertheless, other phases
may exist nearby. In the $\zeta=0$ limit and  $\Delta$ below the studied range, 
the BEG phase diagram 
contains anti-ferromagnetic HS phase separated from the solid HS-LS phase by a narrow
strip of a phase containing both magnetic and HS-LS order. 
For finite $\zeta$ the boundary
between the S and SF provides an interesting possibility
for a stable supersolid phase. Although it was excluded for 2D spinless bosons~\cite{bartouni00,schmid02}
with a simple nn repulsion, the effect of the additional terms in (\ref{eq:boson2})
or the departure from the strong-coupling limit is unexplored. Another interesting
question is the possibility of coexistence of the SF and AF magnetic orders,
observed in the bosonic t-J model with anisotropic exchange~\cite{boninsegni08}.

Our investigation of the Hubbard model in the vicinity of spin-state transition
was motivated by the physics of LaCoO$_3$. While a two-band model ignoring
the electron-lattice coupling is probably too simplistic to describe this 
complicated multi-orbital material, some useful insights are obtained. In particular, the present study
shows that the excitonic condensation is in a broad range of parameters
preferred to the HS-LS order, an order which has been discussed in LaCoO$_3$ context
and treated with first-principles LDA+U method~\cite{knizek09}. The proposal
of excitonic condensation in this material may be tested on the same level of approximation
by introducing the 'excitonic' instead of the standard mean-field decoupling of the
on-site interaction in LDA+U.

\section{\label{sec:conc}Conclusions}
Using dynamical mean-field theory we have performed an unbiased numerical search 
probing all possible particle-hole instabilities of the two-band Hubbard model in the parameter range close to
the spin-state transition. Our main result is the observation of an instability
towards condensation of spinful excitons. Together with the previously reported
solid HS-LS order, these are the only instabilities of the model in the studied parameter range.
We have shown that keeping other parameters fixed
the bandwidths ratio is the control parameter selecting the leading instability, an observation
which has a particularly simple explanation in the strong coupling limit as tuning the 
ration of nn hopping and nn repulsion in a hard-core bosons model.
The strong-coupling mapping onto spinful hard-core bosons with nn interaction provides
a possibility of electronic realization of some exotic phases observed with
cold atoms. 
Comparing the solid HS-LS order and the superfluid excitonic order we find that the former
does not exist in the weak coupling regime and due to its Ising character can be easily suppressed
by geometrical frustration, while the latter exists both in strong and weak coupling limits 
and due to the continuous character can better adapt to geometrical frustration, e.g. by forming 
a 120$^{\circ}$ order on triangular lattice. 
The main implication for real materials is the fact that the excitonic condensation should
be considered a competitor to the HS-LS order in systems close to the spin-state 
transition.

\begin{acknowledgements}
We thank D. Vollhardt, A. Kampf, A. Kauch, P. Nov\'ak, R.~T. Scalettar and J. Otsuki for discussions
and valuable suggestions. We acknowledge the support of Deutsche Forschungsgemeinschaft through FOR1346
and the Grant Agency of the Czech Republic through project 13-25251S.
\end{acknowledgements}

\appendix
\section{Strong coupling parameters}
The parameters of the bosonic model were obtained by second order perturbation theory
in the hopping using
\begin{align}
&\left(H_{\text{eff}} \right)_{\alpha\beta}=\langle \alpha |H|\beta\rangle +
\nonumber\\
&\frac{1}{2}
\sum_i 
\left(
\frac{\langle \alpha |H|i\rangle \langle i |H|\beta\rangle}{E_{\alpha}-E_i}+
\frac{\langle \alpha |H|i\rangle \langle i |H|\beta\rangle}{E_{\beta}-E_i}
\right),
\end{align}
where $|\alpha\rangle$ and $|\beta\rangle$ are the states built from the local
LS and HS states and $|i\rangle$ is everything else. The formula was evaluated in 
Mathematica using the SNEG package~\cite{zitko}.
\begin{widetext}
\begin{align*}
\mu&=\Delta-3J 
+Z\left(t_a^2+t_b^2\right)\left(
\frac{J'^2}{\Delta'^2 \left(U-5J+2 \Delta '\right)}-
\frac{J'^2}{2\Delta ' \left(\Delta '+\Delta \right) \left(U-2J+\Delta '+\Delta \right)}
-\frac{\Delta '+\Delta }{2 \Delta ' \left(U-2J+\Delta '-\Delta \right)}
\right)\\
&+Z\frac{V_1^2+V_2^2}{2}\left(
\frac{J'^4}{\Delta'^2 \left(\Delta '+\Delta \right)^2 \left(U-5J+2 \Delta '+\Delta \right)}
-\frac{2}{U-2J+\Delta '}+\frac{\left(\Delta '+\Delta \right)^2}
{\Delta'^2 \left(U-5J+2 \Delta '-\Delta
   \right)}\right) \\
K_{\parallel}&=\left(t_a^2+t_b^2\right)
\left(
-\frac{J'^2}{\Delta'^2 \left(U-5J+2 \Delta'\right)}
-\frac{J'^2}{(U+J) \Delta' \left(\Delta'+\Delta \right)}
+\frac{J'^2}{\Delta' \left(\Delta'+\Delta \right)\left(U-2J+\Delta'+\Delta\right)} 
\right.\\ 
&+\left.\frac{\Delta'(U+J-\Delta)+\Delta (3J+\Delta)}{(U+J)\Delta'\left(U-2J+\Delta'-\Delta \right)}
\right)
+\frac{V_1^2+V_2^2}{2} \left(
-\frac{J'^4}{\Delta'^2\left(\Delta'+\Delta\right)^2 \left(U-5J+2 \Delta'+\Delta \right)}
+\frac{4}{U-2J+\Delta'}
\right.\\ 
&-\left.
\frac{\left(\Delta'+\Delta \right)^2}{\Delta'^2 \left(U-5J+2 \Delta'-\Delta \right)}
-\frac{2 (U+J)}{(U+J)^2-\Delta^2}
\right)\\
K_{\perp}&=t_a t_b \left(
\frac{J'^2}{\Delta'\left(\Delta'+\Delta\right) \left(U-2J+\Delta'+\Delta \right)}
+\frac{\Delta'+\Delta }{\Delta'\left(U-2J+\Delta'-\Delta\right)}
\right)
+V_1V_2\frac{2J'}{\Delta'\left(U-2J+\Delta '\right)}\\
K_0&=\left(t_a^2+t_b^2\right)\frac{1}{U+J}+\left(V_1^2+V_2^2\right)\frac{U+J}{(U+J)^2-\Delta ^2}\\
K_1&=-t_a t_b 
\frac{2J' \left(U-2J+\Delta'\right)}{(U+J) \Delta'\left(U-5J+2\Delta'\right)}
-V_1V_2\left(
\frac{J'^2 \left(U-2J+\Delta'+\Delta\right)}{\Delta'\left(\Delta'+\Delta\right)(U+J+\Delta )\left(U-5J+2\Delta'+\Delta\right)}
\right.\\ 
&+\left.
\frac{\left(\Delta'+\Delta\right)\left(U-2J+\Delta'-\Delta \right)}{\Delta' (U+J-\Delta)\left(U-5J+2 \Delta'-\Delta\right)}
\right)\\
K_2&=-\frac{V_1t_a+V_2t_b}{2 \sqrt{2} \sqrt{\Delta'\left(\Delta'+\Delta \right)}}
\left(J'
\left(
\frac{1}{U-2J+\Delta'}
+\frac{1}{U-2J+\Delta'+\Delta} 
+\frac{1}{U+J+\Delta}
+\frac{1}{U+J} \right) \right.\\ 
&+\left(\Delta'+\Delta 
\right) 
\left(\left.
\frac{1}{U-2J+\Delta'}
+\frac{1}{U-2J+\Delta'-\Delta }
+\frac{1}{U+J-\Delta}
+\frac{1}{U+J}
\right) 
\right),
\end{align*}
\end{widetext}
where $\Delta'=\sqrt{\Delta^2+J'^2}$. In Hamiltonian (\ref{eq:hubbard}) we did
distinguish between $J$ in $H^{\text{dd}}_{\text{int}}$ and in $H'_{\text{int}}$.
Nevertheless, the above expressions apply to both the models with 
density-density interaction $H^{\text{dd}}_{\text{int}}$ and the full interaction
$H^{\text{dd}}_{\text{int}}+H'_{\text{int}}$ with the provision that
in the density-density case $K_2=0$ and the other expressions are evaluated for $J'=0$.

\section{Mean-Field Decoupling}
Here we show how a mean-field decoupling of the $(U-2J)\sum_{\sigma}n_{a,\sigma}n_{b,-\sigma}$
term in the interaction gives rise to the spontaneous hybridization in the SF phase.
First, we consider the $J'=V_{1,2}=0$ case with degenerate $\chi^S_{\text{OO}}$ and $\chi^A_{\text{OO}}$
modes. Writing the above term as
\begin{equation} 
-\left(U-2J\right)
(a^{\dagger}_{\uparrow}b^{\phantom\dagger}_{\downarrow})(b^{\dagger}_{\downarrow}a^{\phantom\dagger}_{\uparrow})-
\left(U-2J\right)
(a^{\dagger}_{\downarrow}b^{\phantom\dagger}_{\uparrow})(b^{\dagger}_{\uparrow}a^{\phantom\dagger}_{\downarrow})
\end{equation}
we obtain decoupling
\begin{equation}
\phi_1 a^{\dagger}_{\uparrow}b^{\phantom\dagger}_{\downarrow}+
\phi_1^* b^{\dagger}_{\downarrow}a^{\phantom\dagger}_{\uparrow}+
\phi_{-1} a^{\dagger}_{\downarrow}b^{\phantom\dagger}_{\uparrow}+
\phi_{-1}^* b^{\dagger}_{\uparrow}a^{\phantom\dagger}_{\downarrow},
\end{equation}
using complex fields $\phi_1$ and $\phi_{-1}$, which acquire finite
values
\begin{equation}
\phi_1=\phi_x+i\phi_y\sim\langle b^{\dagger}_{\downarrow}a^{\phantom\dagger}_{\uparrow} \rangle, \quad
\phi_{-1}=\phi_x-i\phi_y\sim\langle b^{\dagger}_{\uparrow}a^{\phantom\dagger}_{\downarrow} \rangle
\end{equation}
in the SF phase. 

If the $\chi^S_{\text{OO}}$ and $\chi^A_{\text{OO}}$ are not degenerate the fields
$\phi_1$ and $\phi_{-1}$ are not independent. In this case we use a decoupling
which based on the symmetric and anti-symmetric modes starting from rewriting
the interaction as
\begin{align}
-&\frac{U-2J}{2}
\left(
 a^{\dagger}_{\uparrow}b^{\phantom\dagger}_{\downarrow}+b^{\dagger}_{\uparrow}a^{\phantom\dagger}_{\downarrow}
\right)
\left(
 a^{\dagger}_{\downarrow}b^{\phantom\dagger}_{\uparrow}+b^{\dagger}_{\downarrow}a^{\phantom\dagger}_{\uparrow}
\right)-
\nonumber\\
&\frac{U-2J}{2}\left(
 a^{\dagger}_{\uparrow}b^{\phantom\dagger}_{\downarrow}-b^{\dagger}_{\uparrow}a^{\phantom\dagger}_{\downarrow}
\right)
\left(
b^{\dagger}_{\downarrow}a^{\phantom\dagger}_{\uparrow}
- a^{\dagger}_{\downarrow}b^{\phantom\dagger}_{\uparrow}
\right)
\end{align}
leading to a decoupling
\begin{align}
&\phi_S
\left(
 a^{\dagger}_{\uparrow}b^{\phantom\dagger}_{\downarrow}+b^{\dagger}_{\uparrow}a^{\phantom\dagger}_{\downarrow}
\right)+
\phi_S^*
\left(
 a^{\dagger}_{\downarrow}b^{\phantom\dagger}_{\uparrow}+b^{\dagger}_{\downarrow}a^{\phantom\dagger}_{\uparrow}
\right)
+
\nonumber\\
&\phi_A
\left(
 a^{\dagger}_{\uparrow}b^{\phantom\dagger}_{\downarrow}-b^{\dagger}_{\uparrow}a^{\phantom\dagger}_{\downarrow}
\right)+
\phi_A^*
\left(
b^{\dagger}_{\downarrow}a^{\phantom\dagger}_{\uparrow}-
 a^{\dagger}_{\downarrow}b^{\phantom\dagger}_{\uparrow}
\right)
\end{align}
with
\begin{equation}
\phi_S\sim\langle
 a^{\dagger}_{\downarrow}b^{\phantom\dagger}_{\uparrow}+b^{\dagger}_{\downarrow}a^{\phantom\dagger}_{\uparrow}
\rangle, \quad
\phi_A\sim
\langle
b^{\dagger}_{\downarrow}a^{\phantom\dagger}_{\uparrow}-
 a^{\dagger}_{\downarrow}b^{\phantom\dagger}_{\uparrow}
\rangle.
\end{equation}
Comparing the corresponding terms in $H_{\text{MF}}$ we see that finite $\phi_S$ implies
$\phi_1=\phi_{-1}^*$ and thus real $\phi_x$ and $\phi_y$. Finite $\phi_A$ on the other
hand implies $\phi_1=-\phi_{-1}^*$ and thus imaginary $\phi_x$ and $\phi_y$.

Since the decoupled term appears in both the $SU(2)$ and density-density interactions the above derivations
applies to both cases. In the $SU(2)$ interaction, which includes the spin-flip term, decoupling
in terms of $a^{\dagger}_{\uparrow}b^{\phantom\dagger}_{\uparrow}-a^{\dagger}_{\downarrow}b^{\phantom\dagger}_{\downarrow}$ is possible, which leads to 
the same mean-field equations and gives rise to the $\phi_z$ component
of the order parameter.

\end{document}